\documentstyle[prl,aps,epsf,psfig,twocolumn]{revtex}
\input epsf

\begin{document}

\newcommand{\ksi}{\mbox{$\Xi^0$}}
\newcommand{\aksi}{\mbox{$\overline{\Xi}^0$}}
\newcommand{\ttbs}{\char'134}
\newcommand{\AmS}{{\protect\the\textfont2
  A\kern-.1667em\lower.5ex\hbox{M}\kern-.125emS}}

\title{
{\large \bf 
$\ksi$ and $\aksi$ Polarization Measurements at 800 GeV/c. 
}}

\author{\parindent=0.in
E.~Abouzaid$^{4}$,
A.~Alavi-Harati$^{12}$,
T.~Alexopoulos$^{12,\dagger}$,
M.~Arenton$^{11}$,
A.R.~Barker$^{5,\ddagger}$,
L.~Bellantoni$^7$,
A.~Bellavance$^{9}$,
E.~Blucher$^4$,
G.J.~Bock$^7$,
S.~Bright$^4$,
E.~Cheu$^{1}$,
R.~Coleman$^7$,
M.D.~Corcoran$^{9}$,
B.~Cox$^{11}$,
A.R.~Erwin$^{12}$,
C.O.~Escobar$^3$,
R.~Ford$^7$,
A. Glazov$^4$,
A.~Golossanov$^{11}$,
R.A.~Gomes$^3$,
P. Gouffon$^{10}$,
K.~Hanagaki$^8$,  
Y.B.~Hsiung$^7$,
H.~Huang$^5$,
D.A. Jensen$^7$,
R.~Kessler$^4$,
K. Kotera$^8$,
A.~Ledovskoy$^{11}$,
P.L.~McBride$^7$,
E.~Monnier$^{4,*}$,
K.S.~Nelson$^{11}$,
H.~Nguyen$^7$,
R.~Niclasen$^5$,
H. Ping$^{12}$,
V.~Prasad$^4$, 
X.R. Qi$^7$,
E.J.~Ramberg$^7$, 
R.E.~Ray$^7$,
M.~Ronquest$^{11}$,
T.~Rooker$^{12}$,
E.~Santos$^{10}$,
J.~Shields$^{11}$,
W.~Slater$^2$,
D.E.~Smith$^{11}$,
N.~Solomey$^{4}$,
E.C.~Swallow$^{4,6}$,
P.A.~Toale$^5$,
R.~Tschirhart$^7$, 
C. Velissaris$^{12}$,
Y.W.~Wah$^4$,
J.~Wang$^1$,
H.B.~White$^7$, 
J.~Whitmore$^7$,
M. Wilking$^5$,
B.~Winstein$^4$, 
R.~Winston$^4$, 
E.T. Worcester$^4$,
M.~Worcester$^4$,
T.~Yamanaka$^8$,
E.~D.~Zimmerman$^5$, and
R.F.~Zukanovich$^{10}$
\vspace*{0.1in}

(The KTeV Collaboration) \\

\vspace*{0.1in}
\footnotesize
$^1$ University of Arizona, Tucson, Arizona 85721 \\
$^2$ University of California at Los Angeles, Los Angeles, California 90095 \\
$^3$ Universidade Estadual de Campinas, Campinas, Brazil 13083-970 \\
$^4$ The Enrico Fermi Institute, The University of Chicago, 
Chicago, Illinois 60637 \\
$^5$ University of Colorado, Boulder, Colorado 80309 \\
$^6$ Elmhurst College, Elmhurst, Illinois 60126 \\
$^7$ Fermi National Accelerator Laboratory, Batavia, Illinois 60510 \\
$^8$ Osaka University, Toyonaka, Osaka 560 Japan \\
$^{9}$ Rice University, Houston, Texas 77005 \\
$^{10}$ Universidade de S\~{a}o Paulo, S\~{a}o Paulo, Brazil 05315-970\\
$^{11}$ The Department of Physics and Institute of Nuclear and 
Particle Physics, University of Virginia, Charlottesville, Virginia 22901 \\
$^{12}$ University of Wisconsin, Madison, Wisconsin 53706 \\

\date{\today}

\vspace*{0.1in}

\parbox{14cm}{
The polarization of $\Xi^0$ and 
$\overline{\Xi}^0$ hyperons produced by 800 GeV/c protons on a BeO 
target at a fixed targeting angle of 4.8 mrad
is measured by the KTeV experiment at Fermilab. 
Our result of 9.7\% for $\Xi^0$ polarization shows no significant
energy dependence when compared to a result obtained at 400 GeV/c
production energy and at twice our targeting angle. The polarization
of the $\overline{\Xi}^0$ is measured for the first time and found
to be consistent with zero. We also examine the dependence of polarization
on production $p_t$. 

\vspace*{0.1in}

{\flushleft PACS numbers: 13.30.Ce, 14.20.Jn \hspace*{\fill}}

}

\normalsize

}

\maketitle

\section{Introduction}

Although there is an abundance of data on the 
subject of hyperon polarization in 
hadroproduction, there is still no theory that can correctly 
explain these data.  The discovery of hyperon polarization at Fermilab 
was unexpected~\cite{b1}.
The simplest application of perturbative 
Quantum Chromodynamics (pQCD) assuming massless quarks predicts no 
polarization effects.  
If one considers a theory where
both proton valence and sea 
quarks can be polarized in the collision, then some correlation between SU(6)
wavefunctions and the measured sign and magnitude of polarization is 
possible~\cite{b2}.  But the pattern of polarization for the various hyperons
is far more complex than this theory can handle.  More significantly,
no theory can explain the observation of anti-hyperon polarization
in high energy collisions.  Considering the latter point, 
it is important to measure the polarization for all the anti-hyperons. 
This letter reports on 
measurements of the energy dependence in $\ksi$ polarization and the first 
measurement of polarization in $\aksi$ production by the KTeV 
experiment at Fermilab, using the decay $\Xi^\circ\rightarrow\Lambda\pi^\circ$.

The production and polarization of hyperons and anti-hyperons
have historically been measured as a function
of hyperon momentum ($p$), production angle
($\theta$), beam energy $(p_{beam})$ and target material~\cite{b3,b4}. 
Results are typically reported as a function of the combined variables of
transverse momentum, $p_t$=$p$ $sin\theta$, and longitudinal momentum 
fraction, also called Feynman $x$,  $x_F$=$p/p_{beam}$.
For a fixed angle of hyperon production, as in the data reported here,
the two variables are
proportional to each other.  
We report our polarization results as a function of $p_t$.

Experiments have shown a wide variety of polarization behavior with respect
to $p_t$ and $x_F$[2,4-8]. 
Reference~\cite{b10} reported 
an equal polarization between $\Lambda$ at 800 GeV/c 
with 4.8 mrad targeting and at 400 GeV/c with 9.6 mrad targeting. 
In contrast, measurements of the  
$\Sigma^+$ and $\overline{\Sigma}^-$ polarization~\cite{b7} 
have shown a decrease of polarization
for  800 GeV/c production when compared to 400 GeV/c prodcution.
And finally, an increase in polarization of the $\Xi^-$ as a function 
of the beam energy, 
from 400 GeV/c to 800 GeV/c,
has been measured~\cite{b13}.

There is not as much polarization data for anti-hyperons.
Most are produced with no polarization.
Exceptions are 800 GeV/c
production of 
$\overline{\Xi}^+$ in E756~\cite{b6} 
and $\overline{\Sigma}^-$ in E761~\cite{b7}. 
In these experiments, the anti-hyperons were produced 
with the same magnitude and sign of polarization as their hyperon 
counterparts. 

\section{Beam}

The data discussed in this letter were obtained by the 
KTeV experiment at Fermilab. 
It used an 800 GeV/c proton beam with a 19 second
spill of $\sim 5 \times 10^{12}$ protons once per minute. 
The 1.1 interaction
length (30 cm) BeO target was targeted at a mean downward angle of 
4.80 $\pm$ 0.15 mrad with a rms spread of $\pm$ 0.24 mrad.
Downstream of this target was a set of sweeping magnets used to remove 
charged  particles.  At this targeting angle and beam energy, then
$x_F=0.26p_t (GeV/c)$ 

A right handed KTeV coordinate system was defined with the 
$z$-axis along the hyperon beam momentum $({\bf p_{\Xi}})$ 
and the $y$-axis along the vertical direction. 
For strong production processes polarization should be
normal to the production plane.
Polarization is defined as positive when it is 
along the normal unit vector
$\hat{n} = 
\hat{p}_{p} \times \hat{p}_\Xi$,
where $\hat{p}_{p}$ and $\hat{p}_{\Xi}$ 
are unit vectors along the incident proton beam momentum
and along the produced $\Xi$ momentum respectively. 
The $\hat{n}$ vector is along the negative $\hat{x}$ direction 
of the KTeV coordinate system. 

Located in the target area,
the neutral beam sweeping magnets 
had their
magnetic fields oriented in the vertical $(\hat{y})$ direction.
Their combined field value was set such that 
the $\ksi$ spin was precessed into the $\hat{z}$-direction.
By switching
the polarity of a final rotational 
magnet (RoM), whose field was parallel to
the $\hat{x}$-direction, the $\ksi$ spin was finally precessed 
alternately into the positive and then negative 
$\hat{y}$-axis approximately once a day.  We used the
known value of the $\ksi$ magnetic moment~\cite{b11} to set the
magnet field values.

A series of 
collimators defined two nearly parallel neutral beams
that entered the KTeV apparatus
94~m downstream from the target. The composition of the beams was mainly 
neutrons and $K_L$'s, with a small admixture of $\Lambda$'s and
$\Xi$'s. 

\section{Detector}

The KTeV apparatus consisted of a 
65~m  vacuum
($\sim$$10^{-6}$~Torr) decay region followed by a charged particle 
spectrometer.  This spectrometer consisted of a dipole
analysis magnet (AnM) with two drift
chambers  on either side of it.  
To reduce multiple scattering,  
helium-filled bags occupied the  spaces between 
the drift  chambers.  For the data discussed here, 
the magnetic field imparted a 
$205$~MeV/c
horizontal   momentum   component to   charged   particles.  The AnM
field direction was
reversed approximately once a day.
The momentum  resolution of the spectrometer was 
$\sigma(p)/p\  =   0.38    \%\,  \oplus\,
0.016\%~p\,$(GeV/c), where the sum is in quadrature. 

The charged particle spectrometer was followed by
a ($1.9 \times 1.9$~m$^2$)  electromagnetic 
calorimeter (ECAL),  which consisted
of 3100 pure  CsI crystals. 
After calibration, the ECAL energy  resolution was better than 1\%
for an electron momentum between 2 and 60 GeV/c.
The position resolution was 1~mm. 

Nine photon veto  assemblies  detected  particles leaving  the  fiducial
volume. 
Two scintillator  hodoscopes  in front of the
ECAL were used to  trigger  on  charged particles.  Another scintillator 
plane (hadron-anti), located behind both the  ECAL  and a 10~cm lead
wall, acted as a hadron shower veto. The hodoscopes 
and the ECAL detectors had
two holes, and  the hadron-anti had a single
hole to let the neutral beams pass through without 
interaction. Charged particles passing through these holes were detected 
by $16 \times 16$~cm$^2$ scintillators (hole counters) located along each 
beam line in the hole region just downstream of the hadron-anti. 
Two steel walls, with two sets of hodoscopes, acted as a muon identifier.

Decays of the type 
$\ksi\rightarrow \Lambda \pi^0$($\aksi\rightarrow \overline{\Lambda}\pi^0$) 
produce a high momentum ($>$100~GeV/c) positive(negative) track
proton(antiproton) which remains in or near the neutral beam region.  
In addition
there is a second lower momentum  $\pi^-$($\pi^+$) track
and two 
ECAL energy clusters not associated with tracks that correspond to the
two photons from the $\pi^0$ decay.  
Our trigger was based on these
signatures. 

\section{Data Analysis}

We reconstruct $\Xi^\circ\rightarrow\Lambda\pi^\circ$ decays and the
antihyperon counterpart from both charged track and ECAL cluster quantities.
From the 
energy and position of the photons entering the calorimeter, the 
position of the $\ksi$ vertex along the beam is calculated as
$(\Delta z) = (\delta/m_{\pi^0})\sqrt{E_1 E_2}$,
where $\Delta z$ is the distance of the vertex from the ECAL position,
$E_1$ and $E_2$ are the energies of the photons, $\delta$ is the distance
between the photons at the calorimeter
and $m_\pi^0$ is the known~\cite{b11} $\pi^0$ mass.
The reconstructed proton and $\pi^-$ momentum vectors, 
in the $\ksi$ case, are combined to form the $\Lambda$
momentum vector, which defines its flight path.
The $\Lambda$ flight path is extrapolated   
backwards to the
point where it intersects the $z$-plane of the $\ksi$ decay. 
This intersection yields the $x$ and $y$ coordinates of
the $\ksi$ decay vertex. 
Further details of the KTeV hyperon trigger
and analysis cuts used to reduce background
can be found in  previous papers~\cite{ana1,ana2}.

\begin{figure}[htbp]
\psfig{figure=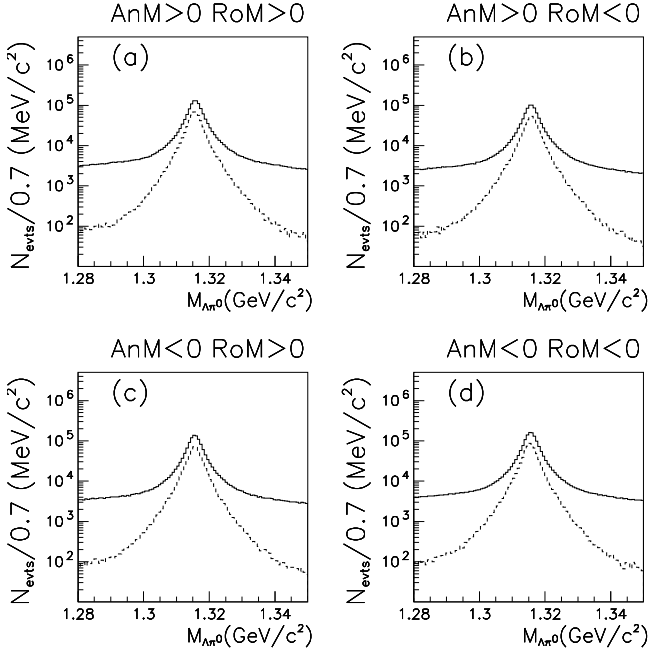,height=3.7in }
\caption{Reconstructed cascade mass from data for various Analysis
(AnM) and
Rotational Magnet (RoM) conditions, before (solid line) and
after (dash line) the application of the
analysis cuts.}
\label{fig:mass_cuts}
\end{figure}

For the selected data,
Fig.~\ref{fig:mass_cuts} shows the effect on the reconstructed 
$\ksi$ mass  after all cuts for various combinations of
analysis magnet and spin rotational  
magnet settings. The remaining background
levels for the $\Lambda \pi^0$ mass peaks are 
less than 0.5\%.

The $\ksi$ or $\aksi$ polarization was determined 
by first splitting our data into two oppositely polarized samples 
(RoM$>0$, RoM$<0$)
and then calculating the direction cosines 
$(\cos \theta_x,\cos \theta_y,\cos \theta_z)$ 
of the $\Lambda$ momentum vector in the $\ksi$ rest frame. For a sample 
of decays where the $\ksi$ has an average polarization $P$, the 
normalized direction cosine distribution ($f_\pm$) is
\begin{equation}
f_\pm(\cos \theta_k) = \frac{dN}{d \cos \theta_k} =  
A(\cos \theta_k)(1 \pm \alpha_{\ksi} P_k cos\theta_k) 
\end{equation} 
\noindent where $k=x,y,z$ and $P_k$ is the $k$-component of polarization $P$
along the $\hat{k}$-axis.
$A(\cos \theta_k)$ is a function that describes  the experimental acceptance
for $\ksi$ decays as a function of the $\Lambda$ direction cosine and 
is a strong function of the analysis magnet (AnM) setting 
for the $cos\theta_x$ distribution. 
$\alpha_{\ksi}$ 
is the known $\ksi$ decay asymmetry 
parameter~\cite{b11}.
The quantity $f_+$ ($f_-$) is proportional to the fraction of the 
up (down) 
precession sample for a given value of $\cos \theta_k$. 

\begin{figure}[htbp]
\psfig{figure=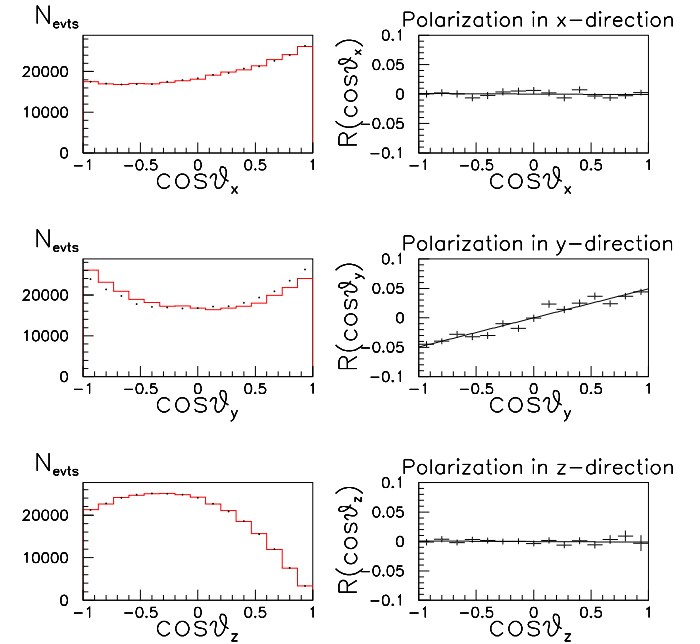,height=3.7in }
\caption{Normalized direction cosine distributions $f_\pm$
(RoM$>0$/RoM$<0$)
in $\hat{x},\hat{y},\hat{z}$
direction
for $\ksi \rightarrow \Lambda \pi^0$ decays, on the left.
Histogram and dots represent the $f_+$  and $f_-$
distributions respectively.
Graphs on the right, show the ratio $R(\cos\theta_k)$.
Error bars are statistical only.}
\label{fig_res1}
\end{figure}

The anti-symmetric ratio:
\begin{equation} \label{eq1}
R(\cos \theta_k) = \frac{(f_+ - f_-)}{(f_+ + f_-)} = 
\alpha_{\ksi} P_k \cos \theta_k
\end{equation}
has a slope with respect to $\cos \theta_k$ which gives the 
asymmetry $\alpha_{\ksi} P_k$, from which the polarization component 
$P_k$ is obtained. 
As long as the acceptance of the detector is factorable and 
does not vary rapidly with time, it cancels out in the ratio. 
Figure~\ref{fig_res1} shows a comparison between the $\Lambda$
direction cosine distributions for the two RoM rotational magnet settings. 
We have combined the data from the two AnM settings 
for each RoM setting in this figure.
As can be seen in the plots in the left column of Fig. 2,
the pairs of distributions are essentially identical 
in the $\hat{x}$ and $\hat{z}$ directions. In the $\hat{y}$ direction, however,
the two distributions are clearly different, showing the effect 
of the $\ksi$ polarization
on the $\Lambda$ decay distribution. 

These effects are even more visible in
the plots of the ratio defined in Eq.~(\ref{eq1})
and are shown in the right column of Fig.~\ref{fig_res1}. Linear fits  
to these graphs extract the polarization components $P_x$, $P_y$, $P_z$. 
The extracted fitted slopes for the $\hat{x}$ and $\hat{z}$ directions 
are consistent with zero ($P_x=P_z=-0.001\pm0.003_{stat}$), 
while the $\hat{y}$ slope shows a clear 
indication of the polarization effect.  
After taking into account 
acceptance differences of the order of $10\%$, using Monte Carlo
generated events~\cite{ana1,ana2}, 
the extracted polarization in the $\hat{y}$-direction
is $P_y = -0.097\pm 0.007_{stat}
\pm 0.013_{sys}$. The systematic error is based on the largest value for
polarization seen in the $\hat{x}$ and $\hat{z}$ directions. We also 
checked that varying analysis cuts by 10\% of their
nominal values had no significant effect on polarization values.  Monte
Carlo reconstructed values for polarization track linearly with input
polarization, at a level much better than the stated systematic error.
The error on $\alpha_{\Xi^\circ}$ is included in the $P_y$ result.

\begin{figure}[htbp]
\psfig{figure=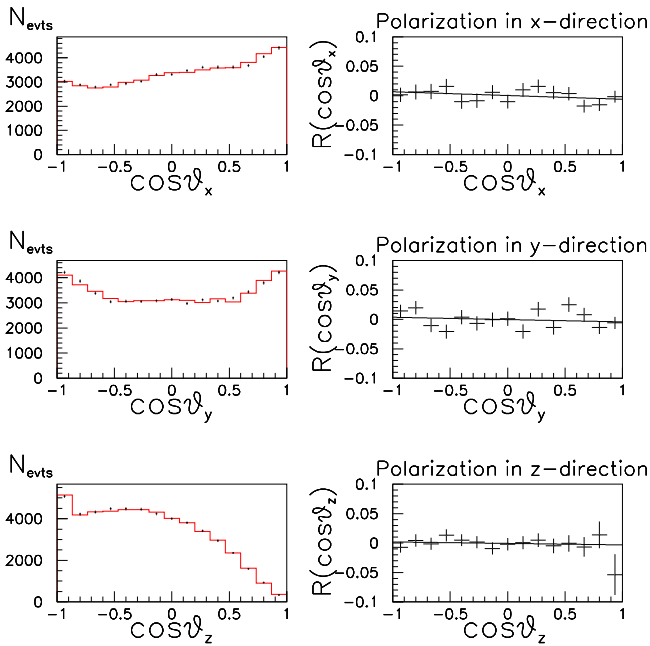,height=3.7in }
\caption{Normalized direction cosine distributions
$f_\pm$ in $\hat{x},\hat{y},\hat{z}$
for $\aksi \rightarrow \overline{\Lambda} \pi^0$ decays, on the left.
Histogram is the $f_+$ and dots represent the $f_-$ distribution.
Graphs on the right, show the ratio $R(\cos\theta_k)$.
Error bars are statistical only.}
\label{fig_res2}
\end{figure}

We carried out a polarization analysis for the
$\aksi$ decay mode as well. 
Fig.~\ref{fig_res2}
shows a comparison between the $\overline{\Lambda}$
direction cosine distributions for the two RoM rotational 
magnet settings.
As can be seen, the pairs of distributions are essentially identical 
in all directions, indicating no 
statistically significant polarization effect for the $\aksi$.
Taking into account the small acceptance effect,
the extracted y-component polarization is $P_y=0.000\pm0.013_{stat} \pm 0.013_{sys}$.
The $P_x$ and $P_z$ results remain
consistent with zero.

\begin{figure}[htbp]
\psfig{figure=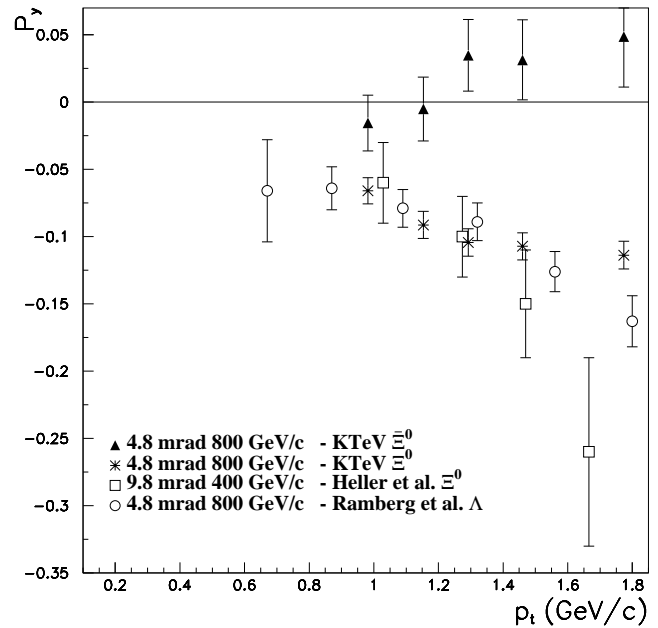,height=3.7in}
\caption{
$\ksi$ and $\aksi$ polarization versus production transverse
momentum $p_t$. For comparison,
$\ksi$ data~\protect\cite{b4} from 400 GeV/c and
$\Lambda$ data~\protect\cite{b10}
from 800 GeV/c are also shown.}
\label{fig_res3}
\end{figure}

To study the detailed momentum dependence of polarization in 
$\ksi$ and $\aksi$ production we divided the data into 
transverse momentum bins.
For each bin, the corresponding polarization   
was calculated. Results are shown in 
Fig.~\ref{fig_res3} as a function of the 
transverse momentum for 
$\ksi$ and $\aksi$. Errors shown are statistical only.  

Comparable data from a past Fermilab experiment
~\cite{b4} have been superimposed on this plot.
Those $\ksi$ data had a targeting angle 
of 9.8 mrad, approximately twice our targeting angle, a proton 
beam energy of 400 GeV, half of our KTeV energy,
and a similar target material (Be).
For a given value of $p_t$, these data 
samples have the same $x_F$  value as the data presented here,
and are therefore directly comparable. No significant change in 
$\ksi$ polarization is seen between the two production 
energies of 400 and 800
GeV. 
Also, the $\Lambda$ polarization from the Fermilab 
experiment E799I~\cite{b10}
is shown, produced with the same targeting angle 
of 4.8 mrad and proton beam 
energy of 800 GeV.  There is a striking similarity 
between $\Lambda$ and $\ksi$
polarization.

\section{Conclusion}

In conclusion, we have measured the polarization of the $\ksi$ and 
$\aksi$ hyperons produced 
by 800 GeV/c protons at a fixed targeting angle 
of 4.8 mrad for the first time. 
Comparing the measured polarization 
values for the $\ksi$ decay mode 
with those determined previously for production 
at 400 GeV/c and a targeting angle of 9.8 mrad, we find 
there is no energy dependence in $\ksi$ production.
We also find no statistically significant polarization for the 
$\aksi$ at 800 GeV/c.  This is in contrast to two previous reports for
other anti-hyperon polarizations~\cite{b7,b6}.
  
\section{Acknowledgments}

We would like to thank L. Pondrom for helpful
discussions. We gratefully acknowledge the support of
the technical staffs of all participating institutions. This work was
supported in part by the US Department of Energy, The National
Science Foundation, The Ministry of Education and Science of 
Japan, 
Funda\c{c}\~{a}o de Amparo a Pesquisa do Estado de S\~{a}o Paulo-FAPESP,
Conselho Nacional de Desenvolvimento Cientifico e Tecnologico-CNPq and
CAPES-Ministerio da Educa\c{c}\~{a}o.

\end{document}